\documentclass[referee]{mn2e}

\usepackage{graphicx}
\usepackage{amsmath}

\title[Misaligned TeV gamma-ray sources in the vicinity of globular clusters]
{Misaligned TeV gamma-ray sources in the vicinity of globular clusters}
\author[W. Bednarek \& T. Sobczak]
{W. Bednarek \& T. Sobczak\\
Department of Astrophysics, The University of Lodz,
ul. Pomorska 149/153, 90-236 Lodz, Poland,\\
bednar@astro.phys.uni.lodz.pl, tmsobczak@uni.lodz.pl}
\begin{document}

\date{Accepted . Received ; in original form }

\pagerange{\pageref{firstpage}--\pageref{lastpage}} \pubyear{2013}

\maketitle

\label{firstpage}

\begin{abstract}
Globular clusters (GCs) contain huge number of low mass stars and also large number of millisecond pulsars (MSPs).
Due to the number of stars, the stellar and MSP winds mix efficiently within the GC. Such mixture of winds leaves GC and 
interacts with the galactic medium creating a bow shock nebula around GC. 
The bow shock nebula is filled with relativistic leptons accelerated in the pulsar magnetospheres and/or wind regions.
We argue that nebulae around globular clusters, immersed in relatively dense medium close to the galactic plane, should have 
complicated morphology due to interaction with the surrounding gas.
Therefore, TeV $\gamma$-ray sources, related to these nebulae, are expected to be misaligned in respect to GC cores,
as observed in the case of GC Ter 5. On the other hand, GCs in low density medium, i.e. far away from the galactic disk, should produce
bow shocks at large distances from the GC cores. The TeV $\gamma$-ray sources around such nebulae are expected to be almost spherical and centred on the GC cores. We perform numerical calculations of the TeV $\gamma$-ray emission produced by leptons escaping from the GC Ter 5. It is shown that TeV $\gamma$-ray source related to Ter 5 should be misaligned in respect to the core of GC as observed by the HESS Collaboration.  

\end{abstract}

\section{Introduction}

Several globular clusters (GCs) have been recently discovered as GeV $\gamma$-ray sources by the {\it Fermi}-LAT telescope
(Abdo et al.~2009a, Abdo et al.~2010, Kong et al.~2010, Tam et al.~2011). The $\gamma$-ray spectra of these GCs
are incredibly similar to those observed from the isolated MSPs in the galactic field (power law  with index 0.7-1.4 with 
exponential cut-off at a few GeV, Abdo et al.~2009b). Therefore, it is very likely that this emission is produced by MSPs within the GCs, as proposed already before this discovery (Harding et al.~2005, Venter et al.~2008, Venter et al.~2009). 

The expected similarity of the radiation processes around classical and millisecond pulsars stimulated observations of
GCs with the modern Cherenkov arrays. Only the upper limits on the TeV $\gamma$-ray flux have been reported based on a relatively 
short observation time of the order of typically several hours (Aharonian et al.~2009, Anderhub et al.~2009, McCutchen et al.~2009, Abramowski et al.~2013). In the most complete study presented in Abramowski et al.~(2013), the upper limits of the TeV flux have been reported for the case of the point like and extended sources. These upper limits are below estimated in this work Inverse Compton flux from leptons injected by the MSPs in the case of some considered GCs.
 
In the case of GC Ter 5, the longer observation ($\sim$ 90 hrs) resulted in the discovery 
of the extended TeV $\gamma$-ray source in the direction of this GC (Abramowski et al.~2011). The $\gamma$-ray spectrum of this source has been measured up to $\sim$20 TeV.
Surprisingly, the centre of the TeV source is shifted from the centre of the GC by the distance corresponding to the dimension of 
the GC (Abramowski et al.~2011). This displacement has put some doubts on the relation of the TeV source to Ter 5. However, no better explanation for its origin is at present known (probability for chance coincidence with PWNa or SNR has been estimated to be low). The identification of this TeV source with Ter 5 seems to be at present the best choice. On the other hand, the {\it Chandra}  observations shows the existence of an extended, likely non-thermal X-ray source centred on the core of Ter 5 (Eger et al.~2010, Clapson et al.~2011). Such X-ray emission could result in the synchrotron process of $\sim$100 TeV electrons in the magnetic field of the order of a few $\mu$G. These leptons could be also responsible for the TeV $\gamma$-ray emission observed from the HESS source towards Ter 5.

GCs with large number of MSPs have been proposed to be potentially new type of TeV $\gamma$-ray sources by Bednarek \& Sitarek~(2007). Specific models for such TeV emissions have been considered in a few papers (e.g. Bednarek \& Sitarek~2007, Venter et al.~2009, Cheng et al.~2010, Kopp et al.~2013, Zajczyk et al. ~2013). These works follow the standard scenario for the $\gamma$-ray production, i.e. the TeV $\gamma$-rays originate in the IC scattering process of low energy radiation (stellar from GC, MBR, or infra-red and optical from the galactic disk) by leptons accelerated by MSPs. Alternative model has been proposed for the origin of TeV $\gamma$-ray source towards Ter 5 by Domainko~(2011). In this scenario $\gamma$-rays appear as a decay products of pions produced in the interactions of hadrons accelerated during explosion of short $\gamma$-ray burst within (or nearby) Ter 5.

In this paper we propose that the TeV $\gamma$-ray sources in the direction of GCs can have complicated morphology due to the interaction of the cumulative winds produced within the GC with surrounding medium. We assume that fast wind, composed of the mixture of the MSP winds and the winds from classical stars, collides with the galactic gas producing a bow shock. The presence of bow shocks around GCs have been proposed in order to explain the diffusive X-ray emission in the vicinity of some GCs (Okada et al.~2007).
Leptons, accelerated around MSPs, are frozen in the wind which propagates preferentially in the direction opposite to the direction of movement of the GC.

\section{Interaction of globular cluster wind with galactic medium}

\begin{figure}
\vskip 7.truecm
\includegraphics{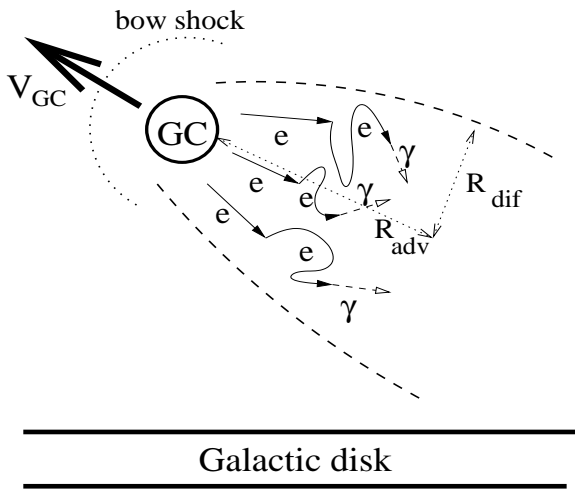}
\caption{Schematic representation (not to scale) of the vicinity of the globular cluster Ter 5 which moves through the interstellar space with velocity $v_{\rm GC}$ in the vicinity of the Galactic Disk. 
The globular cluster wind interacts with the surrounding matter producing a bow shock.
Relativistic leptons, accelerated within the globular cluster, escape preferentially in the directions which are not limited by the bow shock, i.e. opposite to the motion of the cluster. 
The longitudinal structure of the $\gamma$-ray source produced by relativistic leptons (dimension along the direction of motion $R_{\rm adv}$) is determined by the advection process but the perpendicular structure is determined by the diffusion process $R_{\rm dif}$.}
\label{fig1}
\end{figure}

The total energy supplied by the millisecond pulsars into the globular cluster can be estimated from the observed 
GeV $\gamma$-ray luminosity of the globular cluster ($L_\gamma^{\rm GC} = 10^{35}L_{35}$ erg s$^{-1}$) and assuming the efficiency 
of $\gamma$-ray emission by pulsars 
of the order of $\xi = 0.1\xi_{-1}$ (Abdo et al.~2009b), $L_{\rm MSP} = L_\gamma/\xi = 10^{36}L_{35}/\xi_{-1}$ erg s$^{-1}$. This energy is injected into the volume of the globular cluster in the form of relativistic pulsar winds which meet on their way the slow winds dominated by matter from the huge number of low mass stars. Due to the large numbers of stellar winds, both winds should mix effectively producing globular cluster wind containing matter and also relativistic particles.
In fact, different objects within the globular cluster can supply matter in the form of the winds. The contributions could come from normal stars, evaporating companion stars in black widow binary systems and red giants. The contribution of the whole population of the main sequence stars can be estimated on $\sim 10^{-8}$ M$_\odot$yr$^{-1}$ assuming the mass loss rate of specific stars comparable to those observed in the case of the Sun (i.e. $\sim 10^{-14}$ M$_\odot$yr$^{-1}$) and  assuming typical number of stars in GC of the order of $10^6$. The mass loss rate from binary systems of the black widow type (compact binary containing a White Dwarf and MSP)  
is expected to be of the order of $\sim 10^{-10}$ M$_\odot$ yr$^{-1}$. For typical number of black widow binaries in massive GC of the order of 100, the mass loss rate is also of the order of $\sim 10^{-8}$ M$_\odot$yr$^{-1}$.
The red giants seem to be the dominant suppliers of mass into GCs. However, the mass loss rates of red giants in globular clusters are not well known.
Different estimates, from observations of individual stars, ranges between $10^{-9}$ M$_\odot$ yr$^{-1}$ to $3\times 10^{-7}$ M$_\odot$ yr$^{-1}$ (Boyer et al.~2008, Meszaros et al. 2009). Since usually about 100 red giants can be found in specific globular cluster, the mass loss in the range of $10^{-7}$ M$_\odot$ yr$^{-1}$ to $3\times 10^{-5}$ M$_\odot$ yr$^{-1}$ is expected. In our further considerations we estimate the mass loss rate of stars in a globular cluster on
${\dot M}_{GC} = 10^{-6}$ M$_\odot$ yr$^{-1}$. The pulsar winds should mix efficiently with the matter from the stellar winds. As a result the pulsar winds are expected to be slowed down. They still should contain relativistic leptons accelerated in the pulsar winds. The mixed winds should move with the velocity of
\begin{eqnarray}
v_{\rm w} =  \sqrt{2L_{\rm MSP}/{\dot M}_{\rm GC}}\approx 
1.7\times 10^8 ({{L_{35}}\over{M_{-6}\xi_{-1}}})^{1/2}~~~{\rm cm~s^{-1}}
\label{eq1}
\end{eqnarray} 
Therefore, we expect that the globular cluster as a whole create a mixed stellar-pulsar wind which should interact with the medium through which the globular cluster moves with characteristic velocity of the order of $v_{\rm GC} = 3\times 10^7v_7$ cm s$^{-1}$. 
The pressure of this globular cluster wind can be balanced by the ram pressure of the interstellar medium.
Since the globular clusters move around (or even cross) the galactic disk, the density of the medium can change by orders of magnitudes starting from the characteristic value in the galactic disk of the order of 1 cm$^{-3}$ up to $10^{-3}$ cm$^{-3}$, far away from the galactic disk. In our estimation we scale the density of the medium by $n = 0.1n_{-1}$ cm$^{-3}$.
We estimate the distance at which the pressure of the GC wind balances the pressure of the galactic medium on (nv$^2$ = const, see Wilkin~1996),
\begin{eqnarray}
R_{\rm bs} =  \sqrt{{{{\dot M}_{GC}v_{w}}\over{4\pi nv_{\rm GC}^2}}}\approx 
0.8 {{L_{35}M_{-6}^{1/4}}\over{\xi_{-1}^{1/4}n_{-1}^{1/2}v_7}}~~~{\rm pc.}
\label{eq2}
\end{eqnarray} 
\noindent
The radius of the bow shock, estimated for the density of the medium $0.1$ cm$^{-3}$, is comparable to the radius of the GC core 
($\sim 1$ pc). In such case the GC wind should be confined by the bow shock creating a tail in the direction opposite to the direction of movement of the GC. For densities typical for the galactic disk ($\sim 1$ cm$^{-3}$), the mixed pulsar/stellar winds should be very efficiently removed from the GC creating extended tail in the direction opposite to the direction of movement of the GC. However, when the GC is far away from the galactic disk (density $\ll 0.1$ cm$^{-3}$), the shock radius is expected to be an order of magnitude larger than the core radius of the GC. In this last case the quasi-spherical diffusion of the the wind from the GC is expected. 
We conclude that TeV $\gamma$-ray sources with very different morphological structures for the high energy sources related to the winds from the GCs are possible. The basic factor, which determines these structures, is the structure and density of the galactic medium in which the GC is immersed. GCs, which are close to the galactic disk (but outside of it), are expected to produce TeV $\gamma$-ray sources clearly misaligned with the location of the GC. GCs far away from the disk are expected to produce relatively extended spherical sources around GCs. 
In fact, detailed structure of the TeV $\gamma$-ray source can be very complicated depending on the gas density gradients in the vicinity of GC or also on the distribution of MSPs within the GC. 

Relativistic leptons are frozen in the GC wind since the Larmor radius of the TeV leptons is much smaller than the dimension of the 
GC. Therefore, they should follow the morphology of the GC wind. We estimate the characteristic time scale at which leptons can radiate efficiently due to Inverse Compton Scattering of different radiation fields present inside and around the GC. 
The comparison of these energy loss time scales with the dynamical time (determined by the velocity of the wind) gives us an idea about the morphology of the $\gamma$-ray source. The dynamical time scale is
\begin{eqnarray}
\tau_{\rm dyn} = R/v_{\rm w}\approx 1.8\times 10^{10}R_1(M_{-6}\xi_{-1}/L_{35})^{1/2}~~~{\rm s},
\label{eq3}
\end{eqnarray} 
\noindent
where the distance from the GC is $R = 1R_1$ pc. 

The IC energy loss time scales of leptons in the Thomson (T) regime in the Microwave Background Radiation (MBR), with density $0.25$ eV cm$^3$, and the infrared radiation from the galactic disk, with density $1.5$ eV/cm$^{3}$ (e.g. Cheng et al.~2010), can be calculated from standard formula, $\tau_{\rm IC,T} = E_{\rm e}/U_{\rm IC}$, where $U_{\rm IC}$ is the energy loss rate in the T regime.
These time scales are $\tau_{\rm IC,T}^{\rm MBR} = 3.2\times 10^{13}/E_{\rm TeV}$ s and 
$\tau_{\rm IC,T}^{\rm IF} = 6.5\times 10^{12}/E_{\rm TeV}$ s, where the energy of leptons is in TeV.  
Another important (in fact dominating) target radiation for leptons is the optical emission produced by the stars within the GC. However, this radiation field drops with the distance from the GC core. Based on the results presented in Bednarek \& Sitarek~(2007), we approximate the 
energy density of optical photons around GC by applying the simple formula, $U_{\rm opt}\approx 2\times 10^3 L_6 (R_{\rm c}/R)^2$
eV cm$^{-3}$, where $L = 10^6L_6$ L$_\odot$ is the optical luminosity of GC and $R_{\rm c}$ is the core radius of GC.
The energy density of optical photons reach the level of optical background from the galactic disk ($\sim$10 eV cm$^{-3}$, see Cheng et al.~2010) at the distance of $R_{\rm opt}\approx 20L_6^{1/2}R_{\rm c}$. Based on the above approximation, we calculate the IC energy loss time scale for leptons in this optical radiation in the Thomson regime,
$\tau_{\rm IC,T}^{\rm opt,GC} = 4.7\times 10^{9} R^2/(E_{\rm TeV}L_6R_{\rm c}^2)$ s. Note that this formula is valid only for leptons with energies below $\sim$200 GeV. For larger energies the scattering of optical photons occurs in the Klein-Nishina (KN) regime.
We estimate the energy loss time scale in the KN regime by applying the formula for the energy losses of leptons in the Thomson regime but for the lepton energies from the border between the T and KN regimes, i.e for leptons with energies equal to 
$E_{\rm T/KN}\approx m_{\rm e}^2/\varepsilon_{\rm opt}\sim 200$ GeV, where $\varepsilon_{\rm opt}\approx 3k_{\rm B}T_{\rm opt}$, $k_{\rm B}$ is the Boltzmann constant, $m_{\rm e} = 0.511$ MeV is the lepton rest energy, and $T_{\rm opt} = 5000$ K. The density of optical photons within the GC is defined above. Then, $\tau_{\rm IC,KN}^{\rm opt,GC} = 1.2\times 10^{11} R^2E_{\rm TeV}/L_6R_{\rm c}^2)$ s.

By comparing the energy loss time scales of leptons with their dynamical time scale in the GC wind, we obtain the distance at which leptons can lose energy efficiently on scattering of the MBR, $R_{\rm loss}\approx 1.7\times 10^3 (L_{35}/M_{-6}\xi_{-1})^{1/2}/E_{\rm TeV}$ pc,
and the infrared radiation, $R_{\rm loss}\approx 340 (L_{35}/M_{-6}\xi_{-1})^{1/2}/E_{\rm TeV}$ pc. For example, parameters of the GC, considered above, the energy loss distance scale is of the order of $\sim$10 pc for leptons with the maximum energies expected in the MSP winds equal to 40 TeV (Bednarek \& Sitarek~2007). In the case of the optical radiation, this distance scale is
$R_{\rm loss}^{\rm T}\approx 0.26 (L_{35}/M_{-6}\xi_{-1})^{1/2}R^2/(E_{\rm TeV}L_6R_{\rm c}^2)$ pc in the T regime and
$R_{\rm loss}^{\rm KN}\approx 6.7 (L_{35}/M_{-6}\xi_{-1})^{1/2}R^2E_{\rm TeV}/(L_6R_{\rm c}^2)$ pc        
in the KN regime. From the above formulae it becomes clear that low energy leptons are expected to lose significant amount of their energy on the IC scattering of optical photons in the region with the radius of the order of a few pc for the GCs with luminosities of the order of $10^6$L$_\odot$. The above estimates allow us to conclude that leptons frozen in the GC winds should lose significant amount of their energy by scattering mainly the infrared radiation from the galactic disk and the optical radiation produced by stars within the GC.

\section{Morphology of TeV $\gamma$-ray source}
 
The detailed morphology of the TeV $\gamma$-ray source related to GC is determined by the advection process of the GC wind but also by the  diffusion process of leptons in the vicinity of GC. The exact structure of the $\gamma$-ray source depends not only on these two processes but also on the energy dependent losses of leptons on the Inverse Compton process in a few radiation fields, i.e. MBR, infrared radiation from the galactic disk and  optical radiation from the globular cluster itself and also from the galactic disk. Since the energy loss processes in these radiation field are energy dependent (Thomson and Klein-Nishina regimes has to be taken into account) we apply numerical calculations in order to determine the structure of the gamma-ray source. 

We assume that the wind from the GC is very turbulent due to the interaction of many pulsar winds with large number of stellar winds.
Therefore, the Bohm diffusion prescription seems to give good approximation of the diffusion process of leptons in the mixed pulsar stellar winds which escape from GC. The Bohm diffusion coefficient can be expressed by $D_{\rm B} =3\times 10^{25}E_{\rm TeV}/B_{-6}$ cm$^2$ s$^{-1}$, where
$E = 1E_{\rm TeV}$ TeV is the energy of leptons and $B = 10^{-6}B_{-6}$ G is the strength of the magnetic field. The distance scale for the diffusion process can be then expressed by
\begin{eqnarray}
R_{\rm dif} = \sqrt{2D_{\rm B}t}\approx 7.8\times 10^{12}(E_{\rm TeV}t/B_{-6})^{1/2}~~~{\rm cm}.
\label{eq4}
\end{eqnarray} 
\noindent
The dynamical distance scale for the GC wind is $R_{\rm dyn} = v_{\rm w}t\approx 1.7\times 10^8 t(L_{35}/M_{-6}\xi_{-1})^{1/2}$ cm.
From the comparison of these two distance scales, it is found that the advection process of leptons is faster than their diffusion process for the time scale of
\begin{eqnarray}
t_{\rm adv/dif}\approx 2.2\times 10^{9}E_{\rm TeV}M_{-6}\xi_{-1}/(L_{35}B_{-6})~~~{\rm s},
\label{eq5}
\end{eqnarray} 
\noindent
which corresponds to the distance scale $R_{\rm adv/dif}\approx 4\times 10^{17}$ cm for the considered example parameters and leptons with TeV energies.
This distance scale is smaller than the characteristic dimension of the half mass radius of the globular cluster. We conclude that the distribution of relativistic leptons along the direction of the GC wind tail is mainly determined by the dynamics of the wind but the perpendicular distribution of leptons is determined by their diffusion process.

The characteristic dimensions of the $\gamma$-ray source are determined by the characteristic time scale for energy losses of leptons. The shortest energy loss time scale for leptons with the maximum considered energies of 40 TeV is due to the IC scattering of infrared radiation, see $\tau_{\rm IC,T}^{\rm IF}$ defined above.
Then, the longitudinal extend of the $\gamma$-ray source can be estimated on $R_{\rm dyn}\approx 370 (L_{35}/M_{-6}\xi_{-1})^{1/2}/E_{\rm TeV}$ pc and the perpendicular expend on $R_{\rm dif}\approx 6.6/B_{-6}^{1/2}$ pc.
These estimates show that leptons with maximum considered energies mainly radiate in the volume with the characteristic longitudinal extend of the order of ten pc and perpendicular expend of the order of a few pc. The TeV $\gamma$-ray source produced by these leptons  should be additionally shifted from the centre of the globular cluster due to aspherical propagation of the GC wind. Interestingly, estimated above perpendicular extend of the $\gamma$-ray source should not depend on the energy of leptons in the case of scattering of mainly the infrared radiation. This is not the case for leptons with lower energies (close to $\sim$TeV) in which case energy losses can be dominated by scattering of optical radiation from GC.  
 
The image of the $\gamma$-ray source depends on the advection process of leptons from the GC but also on the diffusion process 
(which determines the morphology of the source in perpendicular direction). The diffusion distance scale, in the case of the Bohm
diffusion prescription, depends on lepton energy 'E' and the diffusion time 't' as, $R_{\rm dif}\propto (E\times t)^{1/2}$ (see above).
Since the IC cooling time scale of leptons in the Thomson regime (appropriate for TeV leptons scattering INF photons)
is inversely proportional to the lepton's energy, the diffusion distance scale of leptons does not depend on lepton energy. 
So then, the perpendicular extend of the $\gamma$-ray source in the tail region should not depend on photon energy.
The situation is different in the case of advection
process which determine the propagation process of leptons quasi-radially from the GC. In this case
the advection distance does not depend on lepton energy but depends linearly with advection time (i.e. on the cooling time 
scale of leptons). In such case the efficiency of $\gamma$-ray production, dependent on the cooling time scale of leptons, is 
inversely proportional to lepton energy in T regime  (i.e. when INF photons are scattered). But, In the KN regime (scattering of optical photons), it is proportional to leptons energy (see formulae for estimates of corresponding time scales). 
Therefore, the TeV $\gamma$-ray source should be more extended for $\gamma$-rays with larger energies in the region of GC
but should be less extended with larger energies at the tail region of the nebula.

\section{Misaligned TeV $\gamma$-ray source towards Ter 5}

\begin{figure*}
\vskip 10.truecm
\includegraphics{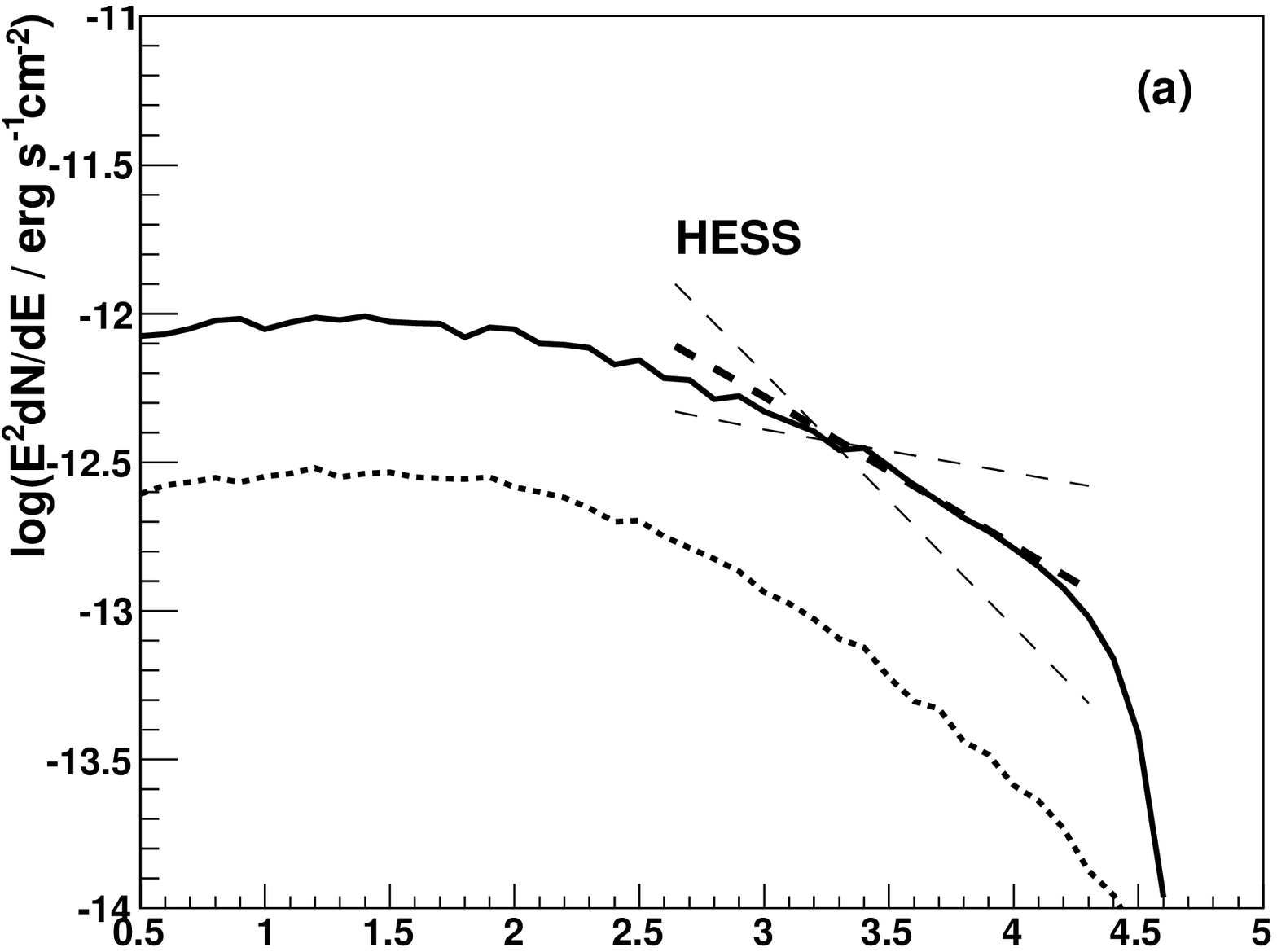}
\includegraphics{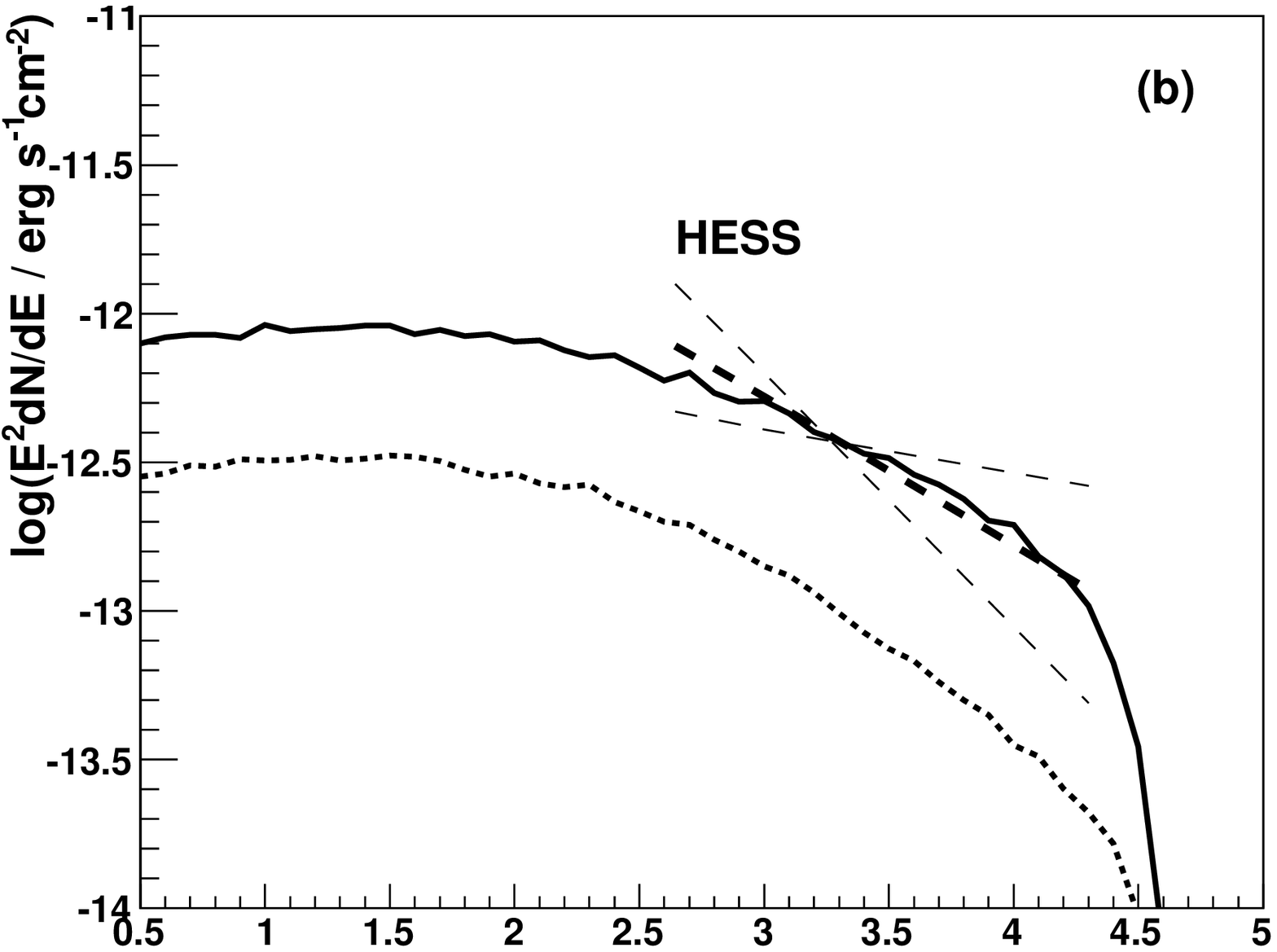}
\includegraphics{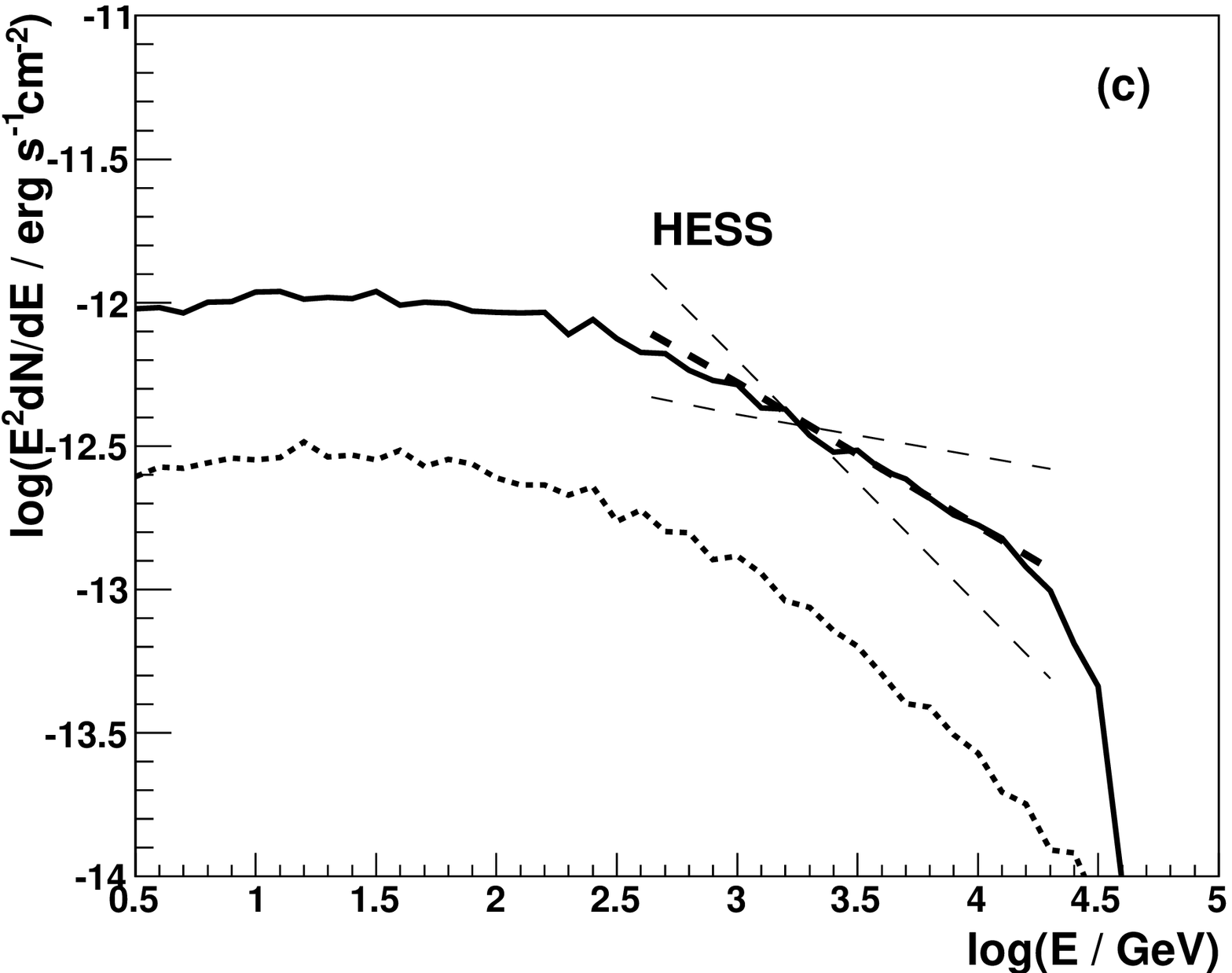}
\includegraphics{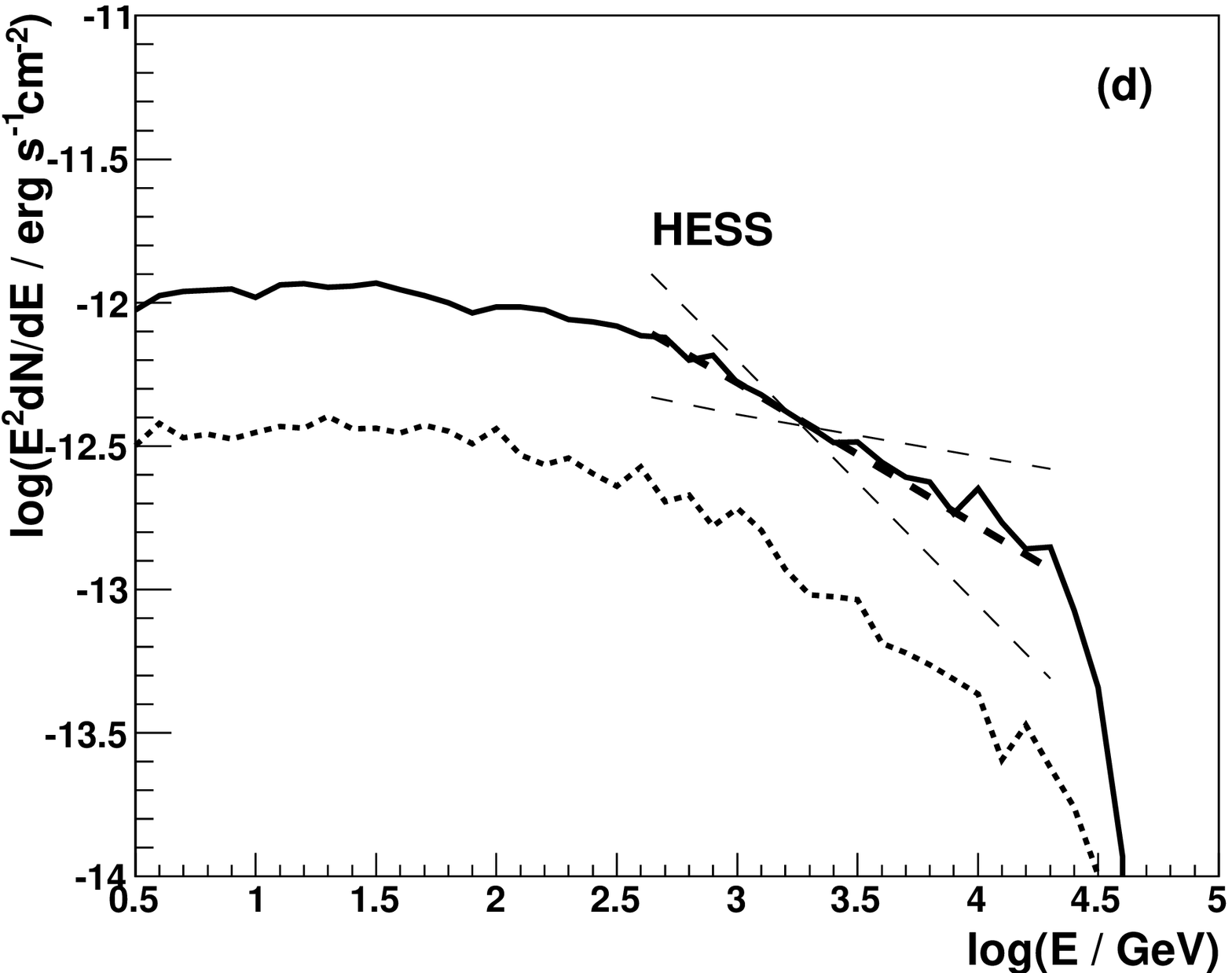}
\caption{Differential IC $\gamma$-ray spectra (SED) produced on both sites of the GC, the bow shock site (dotted curve) and the tail site (solid curve) are compared with the spectrum measured by HESS Collaboration (dashed lines, Abramowski et al.~2011). Two sets of parameters are used
which correspond to the location of the bow shock at the distance of $R_{\rm bs} = 1.1$ pc from the GC and the tail has dimension of $R_{\rm tail} = 16$ pc, corresponding to the angle of the main axis of the bow shock nebula to the observer's line of site equal to $\alpha = 90^\circ$ ((a) and (c)) and for $R_{\rm bs} = 3.5$ pc and $R_{\rm tail} = 32$ pc, corresponding to $\alpha = 30^\circ$ ((b) and (d)). The calculations are performed for two different values of the magnetic field strength in the region of lepton propagation equal to $2\mu$G ((a) and (b)) and $10\mu$G ((c) and (d)).}
\label{fig2}
\end{figure*}

Ter 5 is one of the most luminous GCs, $L_{\rm GC} = 8\times 10^5$ L$_\odot$, with the core radius of $\sim$0.5 pc, taken from Harris~(1996, 2010 edition) see also Lanzoni et al.~(2010). 
It contains 34 radio MSPs. The total number of MSPs is expected to be close to $\sim 200$. This estimation base on the Ter 5 encounter rate and on the observed GeV $\gamma$-ray emission from this cluster (Kong et al.~2010, Abdo et al.~2010).
The cluster contains around 1300 core helium burning horizontal branch (HB) stars (Ferraro et al.~2009). The mass loss rate from these stars is estimated in the range  of $10^{-6}$ M$_\odot$ yr$^{-1}$ to $3\times 10^{-4}$ M$_\odot$ yr$^{-1}$.
The most probable distance to Ter 5 has been estimated on 5.9 kpc (Lanzoni et al.~2010). This GC is located towards the Galactic Centre (galactic coordinates are: l = 3.84$^\circ$ and b = 1.69$^\circ$) within the galactic bulge. Its transversal velocity of Ter 5 has not been measured up till now. It is expected to be rather low (of the order of $\sim$100-150 km s$^{-1}$), given the rotation curve of the Galaxy. Therefore, we can neglect the effects related to movement of this GC on the sky.
   
Ter 5 has been detected by Fermi-LAT with the $\gamma$-ray luminosity $L_\gamma = 2.5\times 10^{35}$ erg s$^{-1}$ (Abdo et al. 2010).
It is the only GC in which direction the TeV $\gamma$-ray source has been detected (Abramowski et al.~2011). The $\gamma$-ray spectrum of this source extends up to $\sim$20 TeV. Interestingly, the source is extended with the size $9.'6\pm 2.'4$ and $1.'8\pm 1.'2$ for the major and minor axis corresponding to the dimensions of $\sim$16$\times$3 pc, for the distance of 5.9 kpc. Moreover, the TeV source is offset from the centre of Ter 5 by $4.'0\pm 1.'9$, corresponding to $\sim$7 pc.

In order to estimate the basic proprieties of the GC wind from Ter 5 and energy loss process of leptons accelerated in the MSP winds, 
we assume the following set of reasonable parameters: $L_{35} = 2.5$, $\xi_{-1} = 1$, $M_{-6} = 10$, $n_{-1} = 10$,  
$v_7 = 1$, $L_6 = 0.8$ and $B_{-6} = 2$. We applied the value for the random magnetic field in the region of lepton propagation slightly lower than the average magnetic field strength in the galactic disk (Ter 5 is at some distance from the galactic disk). In fact, stronger magnetic fields (order of $\sim$10$\mu$G) are expected within a few parsec of the central core of the GC, based on the theoretical estimates (e.g. Bednarek \& Sitarek 2007). Such order of magnetic field strengths are also estimated around Ter 5, based on the observations of the radio emission and assuming minimum energy condition (Clapson et al.~2011). 
No direct measurements of the magnetic field are available and detailed model for possible distribution of the magnetic field within and around the GC is unknown. Therefore, we are forced to perform calculations of the expected TeV $\gamma$-ray emission around Ter 5 in a simple homogeneous model for the magnetic field distribution with the value of 2$\mu$G which seems to be typical for the 
main part of the considered region. For the above parameters, the GC mixed wind velocity is estimated on 
$v_{\rm w}\approx 8.5\times 10^7$ cm s$^{-1}$ (see Eq.~1)
and the distance of the bow shock from the centre of GC on $\sim$1.1 pc (see Eq.~2). The characteristic energy loss distance scale of leptons in the infrared radiation from the galactic disk is estimated on $R_{\rm loss}^{\rm inf}\approx 170/E_{\rm TeV}$ pc and in the optical radiation from the GC itself on $R_{\rm loss}^{\rm T}\approx 0.16 R^2/(E_{\rm TeV}R_{\rm c}^2)$ pc (in the T regime, i.e. for $E_{\rm e}\ll 0.2$ TeV) and $R_{\rm loss}^{\rm KN}\approx 4.2 R^2E_{\rm TeV}/(R_{\rm c}^2)$ pc (in the KN regime). 
These values allows us to conclude that leptons with the largest energies expected in the MSP winds, of the order of $\sim$40 TeV (e.g. Bednarek \& Sitarek~2007), should lose energy on the distance scale of a few pc by scattering infrared radiation still on the border between the T and the KN regimes. However, leptons with lower energies ($\sim$TeV) should mainly lose energy by scattering optical radiation already in the KN regime. These distance scales correspond well to the observed dimensions of the TeV $\gamma$-ray source towards Ter 5.
 
We consider the propagation of leptons from the core of the GC Ter 5 taking into account their advection with the GC wind.
It is assumed that leptons are accelerated with a power law spectrum with spectral index -2.3 up to 40 TeV. The power taken by leptons from the MSP winds is assumed to be a part of the MSP wind power $L_{\rm e^\pm} = \eta L_{\rm MSP}$, where $\eta$ is the acceleration efficiency.
We assumed that the random magnetic field around the GC with the strength of 2$\mu$G.
It is assumed that leptons lose energy on the IC scattering of the infrared and optical radiation from the galactic disk with density 
1.5 eV cm$^{-3}$ and 10 eV cm$^{-3}$ (Cheng et al.~2010) and also on the optical radiation produced by the stellar population in Ter 5. 
The details of the radiation processes of leptons escaping from the GC are discussed in Bednarek \& Sitarek~(2007).
The synchrotron energy losses of leptons are also taken into account as considered in Bednarek~(2012).
The Monte Carlo method is applied in order to determine the distance at which leptons interact with soft photons and the energies of produced $\gamma$-rays. We calculated the TeV $\gamma$-ray spectra expected from the two opposite regions around the GC along the main axis of the bow shock nebula, i,.e  from the region between the GC and the bow shock and from the extended tail region of the GC bow shock nebula. The dimension of the first region is determined by the distance of the bow shock from the GC which depends on the assumed parameters of the GC itself and the surrounding medium. The dimension of the second region is determined by the observed extend of the TeV $\gamma$-ray source. 
We take size of the nebular tail site, corresponding to the maximum extend of the observed TeV $\gamma$-ray source, equal to $R_{\rm tail} = 16$ pc (model a), assuming that the main axis of the bow shock nebula around Ter 5 is perpendicular to the observer's line of sight (inclination angle $\alpha = 90^\circ$). On the other hand, the extend of the emission region is assumed to be equal to 
$R_{\rm tail}/\sin\alpha$ for smaller angles $\alpha$. For example, the real extend of the TeV $\gamma$-ray source is equal to $32$ pc for $\alpha = 30^\circ$ (model b). 
The differential IC $\gamma$-ray spectra from the two regions defined above are compared with the HESS observations of  Ter 5 (Abramowski et al.~2011), in Figs.~2. Since the velocity of the GC in respect to the surrounding medium and the parameters of the medium are not well known in the case of Ter 5, we perform the example calculations for the case of the bow shock located at the distance of 1.1 pc for $n_{-1}^{1/2}v_7 = 3.3$ (model a) and 3.5 pc for $n_{-1}^{1/2}v_7 = 1$ (model b), see Eq.~2. 
The TeV $\gamma$-ray spectra obtained in case of both inclination angles of the bow shock nebula ($\alpha = 90^\circ$ and $30^\circ$) can reproduce well the spectral proprieties of the HESS TeV source (Fig.~2). The GeV $\gamma$-ray emission expected in such model is more than an order of magnitude below the GeV $\gamma$-ray emission observed by Fermi from the core of Ter 5 (Kong et al.~2010). Therefore, the GeV source related to the misaligned TeV $\gamma$-ray source is not expected to be detected by the Fermi-LAT.
Good fit to the observed TeV flux is obtained for the magnetic field strength in the propagation region equal to $2\mu$G and  assuming that the energy taken by relativistic leptons is equal to $\eta = 4.5\%$ of the rotational energy lost by the pulsar for the model (a) (shown in Fig.~2a) and $\eta\sim$3.5$\%$ for the model (b) (Fig.~2b). We also consider the effects due to the stronger magnetic field applying the value equal to $10\mu$G. Stronger magnetic field extracts more efficiently energy from leptons in synchrotron process and also slows down the diffusion process. Good description of the observed TeV $\gamma$-rays spectra is also obtained (see Figs.~2c and 2d). However in this case 
the normalization coefficients are slightly larger $\eta = 5\%$ (for the model (a)) and $\eta = 4.5\%$ (for the model (b)). 
We conclude that the model can explain the spectral features of the TeV source towards Ter 5 for the magnetic field strengths in the range of $\sim$2-10$\mu$G.

\section{Discussion and Conclusion}

We consider the hypothesis that the TeV $\gamma$-ray source, found by the HESS Collaboration in the direction of Ter 5, is in fact related to this globular cluster. We argue that the misalignment between Ter 5 core  and TeV $\gamma$-ray source is due to the aspherical expansion of the GC wind caused by the interaction of the GC wind with the galactic medium.
As a result, a bow shock nebula is formed around Ter 5.
We show that, in the case of GCs immersed in a relatively dense environment, the nebulae around GC should be highly anisotropic. The mixed winds from the MSPs and stars within GC should be confined from the site of the bow shock. On the other hand, mixed winds are expected to expand freely in the direction of nebular tail. Relativistic leptons, accelerated in the winds of MSPs, are confined in the GC wind and interact with the optical radiation from the stellar population within GC, optical and infrared radiation from the galactic disk, and the Microwave Background Radiation producing TeV $\gamma$-rays. We show that the bow shock nebulae around GCs are strongly anisotropic in the case of GCs close to the galactic disk were the GCs are immersed in dense medium. As an example, we consider in more detail the case of GC Ter 5 which is located in the direction of the Galactic Centre within the Galactic Bulge. We have performed calculations of the TeV $\gamma$-ray emission produced by leptons in the GC wind assuming likely parameters characterising GC Ter 5 and its surrounding medium. The $\gamma$-ray emission from two regions, the bow shock site and the tail site, have been compared with the HESS observations of Ter 5. We conclude that the TeV $\gamma$-ray source should be clearly asymmetric in respect to the GC core.
However, in the case of GCs moving through the galactic medium with a relatively low density (far away from the galactic disk), the bow shocks form at relatively large distances from the GCs. As a result, the GC nebula is more isotropic. The TeV $\gamma$-ray sources, related to such nebulae, are expected to be quite spherical and centred on the GC cores. Therefore, we predict that misaligned TeV $\gamma$-ray sources appear only in the vicinity of GCs close to the galactic plane such as Ter 5.
We conclude that the constraints on the IC leptonic models of the type presented in Abramowski et al.~(2013) can depend not only
on differences in the acceleration efficiency of leptons in specific GCs and the strengths of their magnetic field but also on
the environment of the GC which can be responsible for the complicated morphology of the TeV $\gamma$-ray source.
In order to test the expected relation of the morphology of the TeV $\gamma$-ray source with the features of the GC environment, much deeper observations of the population of GCs (at different distances for the galactic disk), than recently presented in Abramowski et al.~(2013), should be performed.

The magnetic field in the propagation region of leptons in the range $2-10\mu$G, expected within and around
GC (Bednarek \& Sitarek~2007, Clapson et al.~2011), does not effect significantly the $\gamma$-ray spectra produced in terms of this model. This is due to the assumption that dynamics of leptons is determined by the advection process which is independent on 
the magnetic field. The energy losses of leptons are still dominated by the IC process and the synchrotron energy losses play minor role.
 
In terms of discussed model we do not predict the existence of a diffusive keV synchrotron emission
related to the misaligned TeV $\gamma$-ray source towards Ter 5. The maximum energies of synchrotron 
photons produced by 40 TeV leptons in the magnetic field of $2\mu$G are of the order of
$\varepsilon_{\rm syn}\approx m_{\rm e}c^2 (B/B_{\rm cr})\gamma_{\rm e}^2\sim 0.1$ keV, where $B_{\rm cr} =4.4\times 10^{13}$ G 
is the critical magnetic field strength and $\gamma_{\rm e}$ is the Lorentz factor of leptons.
They are clearly below X-ray range. However, in the core of GC much stronger magnetic field might be present,
see for theoretical estimates e.g. Bednarek \& Sitarek (2007) and experimental estimates (Clapson et al.~2011).
If the magnetic fields in the central part of GCs are clearly stronger that considered in this paper, 
then the existence of diffusive X-ray emission around the core of Ter 5 could be consistent with the scenario discussed
in this paper. However, detailed structure of the magnetic field within and around Ter 5 is unknown 
which make more realistic calculations impossible on the present stage.

\section*{Acknowledgments}
This work is supported by the grant through the Polish Narodowe Centrum Nauki No. 2011/01/B/ST9/00411.


\label{lastpage}
\end{document}